\documentclass[a4paper]{article}

\usepackage{INTERSPEECH2020}
\usepackage{mathrsfs}
\usepackage{diagbox}
\usepackage{multirow}
\usepackage{subfigure}
\usepackage{threeparttable}

\title{Spike-Triggered Non-Autoregressive Transformer for \\ End-to-End Speech Recognition}
\name{Zhengkun Tian$^{1,2}$, Jiangyan Yi$^{1}$, Jianhua Tao$^{1,2,3}$, Ye Bai$^{1,2}$, Shuai Zhang$^{1,2}$, Zhengqi Wen$^{1}$}
\address{
	$^1$NLPR, Institute of Automation, Chinese Academy of Sciences, Beijing, China\\
	$^2$School of Artificial Intelligence, University of Chinese Academy of Sciences, Beijing, China\\
	$^3$ CAS Center for Excellence in Brain Science and Intelligence Technology, Beijing, China}
\email{\{zhengkun.tian, jiangyan.yi, jhtao, ye.bai, shuai.zhang, zqwen\}@nlpr.ia.ac.cn}
\begin{document}

\maketitle
\begin{abstract}
Non-autoregressive transformer models have achieved extremely fast inference speed and comparable performance with autoregressive sequence-to-sequence models in neural machine translation. Most of the non-autoregressive transformers decode the target sequence from a predefined-length mask sequence. If the predefined length is too long, it will cause a lot of redundant calculations. If the predefined length is shorter than the length of the target sequence, it will hurt the performance of the model. To address this problem and improve the inference speed, we propose a spike-triggered non-autoregressive transformer model for end-to-end speech recognition, which introduces a CTC module to predict the length of the target sequence and accelerate the convergence. All the experiments are conducted on a public Chinese mandarin dataset AISHELL-1. The results show that the proposed model can accurately predict the length of the target sequence and achieve a competitive performance with the advanced transformers. What's more, the model even achieves a real-time factor of 0.0056, which exceeds all mainstream speech recognition models.
 
\end{abstract}

\section{Introduction}


Although autoregressive models have achieved great success in various NLP tasks and speech recognition\cite{bahdanau2014neural, chorowski2015attention, chan2016listen, vaswani2017attention, kim2017joint, dong2018speech}, the autoregressive characteristics result in a large latency during the inference process \cite{lee2018deterministic}. Most of the attention-based sequence-to-sequence models generate the target sequence in an autoregressive fashion. These models  predict the next token conditioned on the previously generated tokens and the source state sequence. By contrast, the non-autoregressive model gets rid of temporal dependency and able to perform parallel computing, greatly improving the speed of inference.

Non-autoregressive transformers (NAT) have achieved comparable results with autoregressive models in neural machine translation and speech recognition \cite{lee2018deterministic, gu2017non, ma2019flowseq, wang2019non, chen2019non, libovicky2018end, moritz2019triggered}. Different from the autoregressive sequence-to-sequence model, the NAT takes a fixed-length mask sequence as input to predict target sequence. The setting of this predefined length is very important. If the length is shorter than the actual length, this will cause many errors of deletion. On the contrary, a longer length will cause the model to generate duplicate tokens and consume additional calculations. To our best knowledge, there are three ways to estimate the length of the target sequence. Firstly, some works introduce a neural network module behind the encoder to predict the target length \cite{lee2018deterministic, gu2017non, ma2019flowseq}. These method cannot guarantee the accuracy of the predicted lengths. During inference, it is necessary to sample different lengths to select the optimal sequence. Secondly, \cite{wang2019non, chen2019non} set an empirical(or maximum) length based on the length of the source sequence. To guarantee the performance of the model, the length is often much longer than the actual length of the target sequence. It will result in extra calculation cost and affect the inference speed. Thirdly, \cite{libovicky2018end} utilizes the CTC loss function instead of the cross entropy to optimize the model, which makes the model generate tokens without calculating the length of the target sequence. However, the characteristics of CTC will cause the model to generate some duplicate tokens and a large number of blanks during inference, and it does not accelerate the inference speed.

\vspace{-2pt}
For speech recognition, the number of valid characters or words contained in a piece of speech is affected by various factors such as the speaker's speech rate, silence, and noise. It is unreasonable to set a fixed length only according to the duration of the audio. To estimate the length of the target sequence accurately and accelerate the inference speech, we propose a spike-triggered non-autoregressive transformer (ST-NAT) for end-to-end speech recognition, which introduces a CTC module to predict the length of the target sequence and accelerate the convergence. The CTC loss plays three important roles in our proposed model. Firstly, ST-NAT utilizes the CTC module to predict the length of target sequences. The CTC module can generate spike-like label posterior probabilities. The number of spikes accurately reflects the length of the target sequence \cite{ma2019flowseq, moritz2019streaming}. During inference, the ST-NAT can count the number of spikes to avoid redundant calculations. Secondly, the ST-NAT adopts the encode states corresponding to the positions of spikes as the input of the decoder. We assume that the triggered encode state sequence can obtain more prior information than the mask sequence, which may able to improve the performance of the model. Thirdly, the ST-NAT adapts the CTC loss as an auxiliary loss to speed up training and convergence \cite{kim2017joint}. Additionally, a non-autoregressive transformer cannot model the inter-dependencies between the outputs. Therefore, we improve the model performance by integrating the output probabilities predicted by the ST-NAT and a neural language model. All experiments are conducted on a public Chinese mandarin dataset AISHELL-1. The results show that the ST-NAT can predict the length of the target sequence accurately and achieve comparable performance with the most advanced end-to-end models. The probability of missing words or characters is less than 2\%. What's more, the model even achieves a real-time factor (RTF) of 0.0056, which exceeds all mainstream speech recognition models.

\vspace{-2pt}
The remainder of this paper is organized as follows. Section 2 describes our proposed triggered non-autoregressive transformers. Section 3 presents our experimental setup and results. The conclusions and future work will be given in Section 4.

\section{Spike-Triggered Non-Autoregressive Transformer}

\subsection{Model Architecture}

The spike-trigger non-autoregressive transformer consists of an encoder, a decoder, and a CTC module, as depicted in Fig.1. Both encoder and decoder are composed of multi-head attention layers and feed-forward layers \cite{vaswani2017attention}, which is similar to the speech transformer \cite{dong2018speech}. 

As shown in Fig.1, we put a 2D convolution front end at the bottom of the encoder to process the input speech feature sequences simply, including dimension transformation (from 40 to 320), time-axis down-sampling, and adding sine-cosine positional information.

Multi-head attention (MHA) layer allows the model to focus on the information from different positions. Each head $h_i$ is a complete self-attention component. $Q$, $K$ and $V$ represent queries, keys and values respectively. $d_k$ is the dimension of keys. $W^Q\in\mathbb{R}^{d_m\times{d_q}}$, $W^K \in \mathbb{R}^{d_m\times{d_k}}$, $W^V\in\mathbb{R}^{d_m\times{d_v}}$ and $W^O\in\mathbb{R}^{d_m\times{d_m}}$ are projection parameter matrices. 
\begin{equation}
\label{eq:self-attention}
\text{SelfAttn}(Q,K,V)=\text{softmax}(\frac{QK^T}{\sqrt{d_k}})V
\end{equation}
\begin{equation}
\begin{split}
\text{MultiHead}(Q,K,V)&=\text{Concat}(h_1,h_2,...h_{n_h})W^O \\
\text{where } h_i =& \text{SelfAttn}(QW_i^Q,KW_i^K,VW_i^V)
\end{split}
\end{equation}

Feed-forward network (FFN) contains two linear layers and a gated liner unit (GLU) \cite{dauphin2017language} activation function \cite{Tian2019, fan2019speaker}.
\begin{equation}
FFN(x)=\text{GLU}(xW_1+b_1)W_2+b_2
\end{equation}
where parameters $W_1 \in \mathbb{R}^{d_m\times{2d_{ff}}}$, $W_2\in \mathbb{R}^{d_{ff}\times{d_{m}}}$, $b_1\in\mathbb{R}^{d_{m}}$ and $b_2\in\mathbb{R}^{d_{m}}$ are learnable. 

The sine and consine positional embedding proposed by \cite{vaswani2017attention} are applied for all the experiments in this paper. Besides, the model also apply residual connection and layer normalization.

The ST-NAT introduces a CTC module to predict the length of the target sequence and accelerate the convergence. The CTC module only consists of a linear project layer. Most non-autoregressive transformer model adopts a fixed-length sequence filled with '$\langle\textit{MASK}\rangle$' as the input of the decoder. These sequences don't contain any useful information. In fact, the CTC spike is usually located in the range of one specific word. Therefore, the ST-NAT utilizes the encoded states corresponding to the CTC spike as the input of the decoder. We assume that the triggered encode state sequence contains some prior information on the target words, which makes the decoding process more purposeful than guessing from the empty sequence.

\begin{figure}[t]
	\centering
	\label{fig:whole}
	\includegraphics[width=1.0\linewidth]{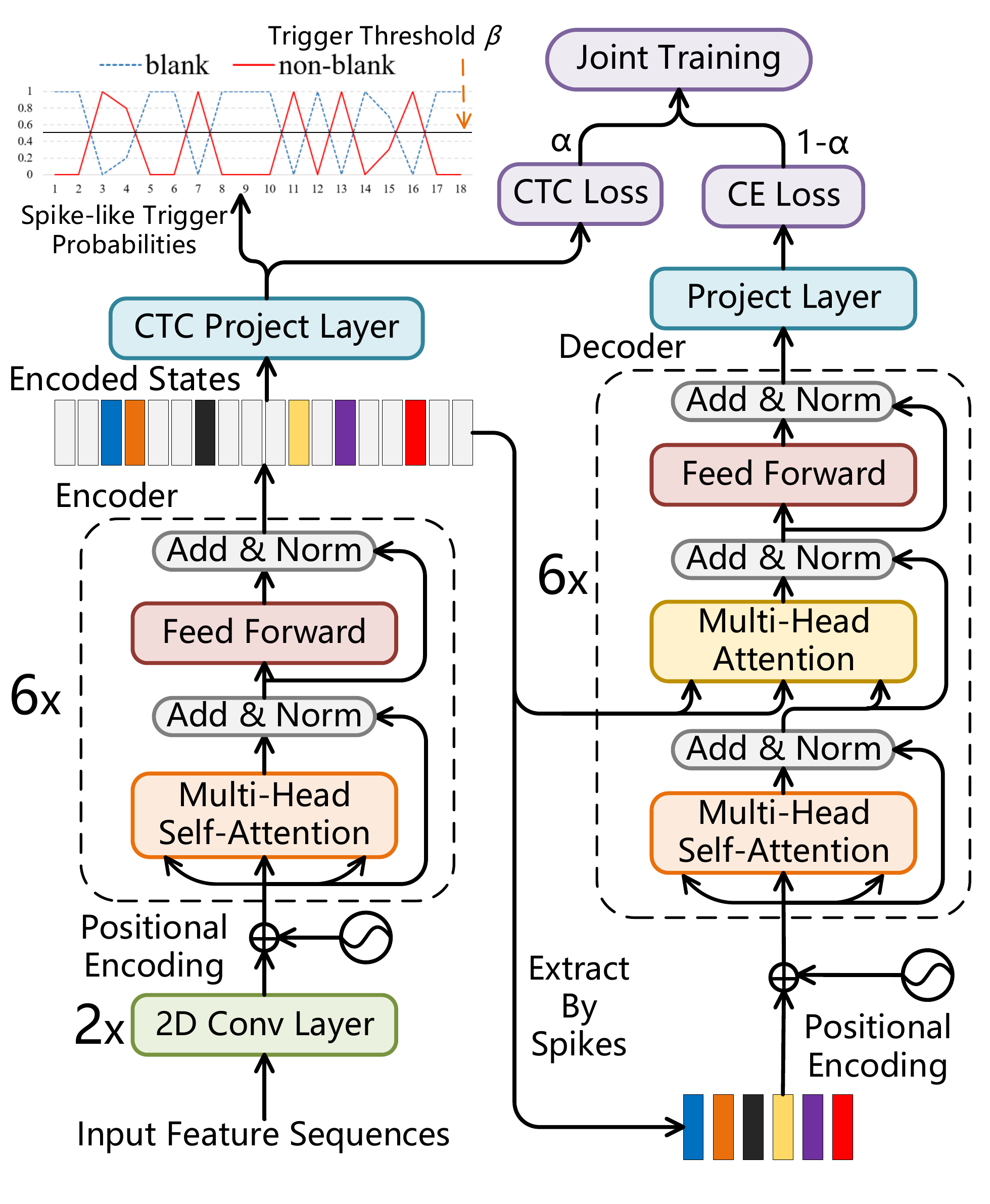}
	\caption{The spike-triggered non-autoregressive transformer has three components, an encoder, a decoder, and a CTC module. The encoder processes the input feature sequences into encoded states sequence. The CTC module computes spike-like posteriors from encoded states. And then the decoder extract encoded states sequence from the corresponding position of spikes as the input. The whole system is trained jointly.}
\end{figure}

\subsection{Training}

It is very important to predict the length of the target sequence accurately. When the predicted length $T^\prime$ is shorter than the target length $T$, there is no doubt that the generated sequence will miss many words or characters, which means that it causes many deletion errors. Instead, the predicted length $T^\prime$ is longer than the target length of $T$, it will cost extra calculation and even generate many duplicate tokens. The ST-NAT can predict the length of the target sequence accurately, through counts the number of spikes produced by the CTC module. When the probability that the CTC module generates a non-blank token is greater than the trigger threshold $\beta$, the corresponding trigger position is recorded. This process can be described as follows.
\begin{equation}
POS(i)= 
\left\{ 
\begin{array}{lr}
triggerd, & 1 - p_b \ge \beta  \\
ignored, & 1 - p_b < \beta
\end{array}
\right.
\end{equation}
Where $POS(i)$ means the $i$-th position of the encoder output states. $p_b$ is the blank probability predicted by the CTC module. Then the probability of non-blank can be expressed as $1 - p_b$. The ST-NAT also inserts an end-of-sentence token '$\langle\textit{EOS}\rangle$' into the target sequence to guarantee the model still able to generate a correct sequence, when the predicted length $T^\prime$ is larger than the target length $T$.


Furthermore, it has been widely proved that the CTC loss function \cite{graves2006connectionist} is effective to assist the model to accelerate the training and convergence \cite{kim2017joint}. It is difficult to train a non-autoregressive model from scratch. Therefore, we use the CTC loss as an auxiliary loss function to optimize the model. 
\begin{equation}
\mathcal{L} = 
\left\{ 
\begin{array}{lr}
\alpha\mathcal{L}_{CTC} + (1 - \alpha)\mathcal{L}_{CE}, & T^\prime \ge T \\
\mathcal{L}_{CTC}, & T^\prime < T
\end{array}
\right.
\end{equation}
where $\mathcal{L}_{CE}$ is the cross entropy loss \cite{de2005tutorial} and $\mathcal{L}_{CTC}$ is the CTC loss. $\alpha$ is the weight of CTC in joint loss function. $T^\prime$ is the predicted target length. $T$ is the real target length. If $T^\prime$ is smaller than $T$, the ST-NAT only utilizes the CTC loss optimize the encoder. Thanks to the CTC module, the ST-NAT can be trained from scratch and without any pre-training or other tricks.

\subsection{Inference}
During inference, we just select the token which has the highest probability at each position. Generating the token '$\langle\textit{EOS}\rangle$' or the last word in the sequence means the end of the decoding process.

Non-autoregressive model cannot model the temporal dependencies between the output labels. This largely prevents the improvement of model performance. We also introduce a transformer-based language model into decoding process. Neural language model makes up the weakness of non-autoregressive model. The joint decoding process can be described as 
\begin{equation}
\begin{array}{lr}
\hat{y} = \arg \underset{y}{\max}(logP(y|x)+\lambda logP_{LM}(y|x))
\end{array}
\end{equation}
where $\hat{y}$ is the predict sequence. And $P_{LM}(y|x)$ is the probability of language model. $\lambda$ is the weight of the language model probabilities.

\section{Experiments and Results}

\subsection{Dataset}
In this work, all experiments are conducted on a public Mandarin speech corpus AISHELL-1\footnote{http://www.openslr.org/13/}. The training set contains about 150 hours of speech (120,098 utterances) recorded by 340 speakers. The development set contains about 20 hours (14,326 utterances) recorded by 40 speakers. And about 10 hours (7,176 utterances / 36109 seconds) of speech is used as test set. The speakers of different sets are not overlapped.

\subsection{Experiment Setup}

For all experiments, we use 40-dimensional FBANK features computed on a 25ms window with a 10ms shift. We chose 4233 characters (including a padding symbol '$\langle\textit{PAD}\rangle$' , an unknown symbol '$\langle\textit{UNK}\rangle$' and  an end-of-sentence symbol '$\langle\textit{EOS}\rangle$') as model units.

Our proposed model and baseline models are built on OpenTransformer\footnote{https://github.com/ZhengkunTian/OpenTransformer}. The ST-NAT model consists of 6 encoder blocks and 6 decoder blocks. There are 4 heads in multi-head attention. The 2D convolution front end utilizes two-layer time-axis CNN with ReLU activation, stride size 2, channels 320, and kernel size 3. The output size of the multi-head attention and the feed-forward layers are 320.  We adopt an Adam optimizer with warmup steps 12000 and the learning rate scheduler reported in \cite{vaswani2017attention}. After 80 epochs, we average the parameters saved in the last 20 epochs. We also use the time mask and frequency mask method proposed in \cite{park2019specaugment} for the baseline transformer, SAN-CTC, and all non-autoregressive models. During inference, we use a beam search with a width of 5 for the baseline Transformer model, SAN-CTC model and the ST-NAT with language model.

We use the character error rate (CER) to evaluate the performance of different models. For evaluating the inference speed of different models, we decode utterances one by one to compute real-time factor (RTF) on the test set. The RTF is the time taken to decode one second of speech. All experiments are conducted on a GeForce GTX TITAN X 12G GPU.

\subsection{Results}

\subsubsection{Explore the effects of different weights and trigger thresholds.}
We train the ST-NAT model with different CTC weights $\alpha$ and trigger thresholds $\beta$ from scratch. As shown in Table.\ref{tab:trigger}, the trigger NAT model with CTC weight 0.7 and trigger threshold 0.3 can achieve a CER of 7.66\% on test. At the same threshold, the trigger NAT with weight 0.6 can achieve the best performance on development set. The CTC weights and trigger thresholds affect the performance of the model in different aspects. The CTC weights $\alpha$ are used to balance the performance of CTC trigger module and decoder. However, the trigger threshold $\beta$ are used to determine how many encoder states are triggered. Both weights and thresholds play important roles in the performance of the models.

\begin{table}[t]
	\caption{Comparison of the model with different CTC weights $\alpha$ and trigger thresholds $\beta$. We evaluate the CER(\%) on development and test set, respectively.}
	\label{tab:trigger}
	\centering
	\begin{tabular}{c|c|c|c|c}
		\bottomrule
		\textbf{CTC} & \multicolumn{4}{c}{\textbf{Trigger Threshold} $\bm{\beta}$} \\
		\cline{2-5}
		\textbf{Weight} $\bm{\alpha}$&0.1&0.3&0.5&0.7\\
		\hline
		0.1&7.67/8.66&7.56/8.50&7.66/8/45&7.56/8.50\\
		0.3&7.37/8.14&7.25/8.12&7.26/8.19&7.21/8.01\\
		0.5&7.06/7.97&7.10/7.88&7.38/8.14&7.30/8.12\\
		0.6&7.06/7.88&\textbf{6.88}/7.67&7.05/7.77&7.01/7.70\\
		0.7&7.26/8.05&6.91/\textbf{7.66}&7.03/7.87&7.39/8.02\\
		\bottomrule
	\end{tabular}
\end{table}

\begin{table}[t]
	\caption{Comparison of the effects of different trigger threshold on the inference speed. We record the time that the ST-NAT spends on decoding test set and calculate the real-time factor.}
	\label{tab:rtf}
	\centering
	\begin{tabular}{c|cccc}
		\toprule
		\textbf{Threshold} $\bm{\beta}$ & \textbf{0.1} & \textbf{0.3} & \textbf{0.5} & \textbf{0.7} \\
		\hline
		Performance & 7.88 & 7.67 & 7.77 & 7.70 \\
		Seconds & 212.04 & 202.59 & 200.62 & 198.44  \\
		RTF & 0.0059 & 0.0056 & 0.0055 & 0.0054 \\
		\bottomrule
	\end{tabular}
\end{table}

\subsubsection{Explore the effects of different trigger thresholds on the inference speed.}

We evaluate our ST-NAT with different trigger thresholds on the inference speed. All the ST-NAT models are trained with a CTC weight of 0.6. It is obvious from the Table.\ref{tab:rtf} that the larger the threshold, the faster the model decode an utterance. When the trigger threshold is 0.7, the model achieves an RTF of 0.0054. It also means the model only has a latency of nearly 20 milliseconds. However, a large threshold does not mean that the model can achieve the best performance. A large trigger threshold might cause the predicted length generated by the CTC trigger to be shorter than the target length, which in turn will hurt the performance of the model. Fortunately, different trigger thresholds have only little effect on the speed of inference, which can even be ignored.

\subsubsection{Analysis on trigger mechanism.}

We analyze the trigger non-autoregressive transformer from the following two perspectives. 

On the one hand, we explore the relationship between the predicted length by the model and the target length, as show in Fig.2. The histogram record the difference between the target length and the predicted length. When the value is less than or equal to zero, it means that the predicted length is less than or equal to the target length. This will not cause irreversible effects. The decoder is still able to predict a token  at the end of sentence. We can find that the vast majority of predicted length have no any errors. What's more, the probability of missing words or characters is even less than 2\%. Even for the most of weights (0.3, 0.5 and 0.7), the maximum predict error does not exceed 4. Therefore, we conclude that the CTC model can predict the length of the target sequence approximately accurately. However, if the value is larger than zero, the model will miss some words permanently. We can fix this problem by adding a padding bias to the predicted length.

On the other hand, Fig.\ref{fig:ctc} shows the relationship between the trigger position and the word pronunciation boundary. There is no triggered spike in the range of silence. Within the scope of the last pronounced word, there are two triggered spikes. Because we also take an end-of-sentence token into consideration during training. It's obvious that each spike is within the boundary of the word. Therefore, our assumption, that the triggered encode state sequence contains more prior information on the target sequence, is reasonable. It's obvious from Fig.\ref{fig:attn} that the ST-NAT model can make the target sequence better aligned to the encoded states sequence. What's more, the center of the alignment position almost coincides with the trigger position, which again verifies our assumption.

\begin{figure}[t]
	\centering
	\label{fig:len}
	\includegraphics[width=\linewidth]{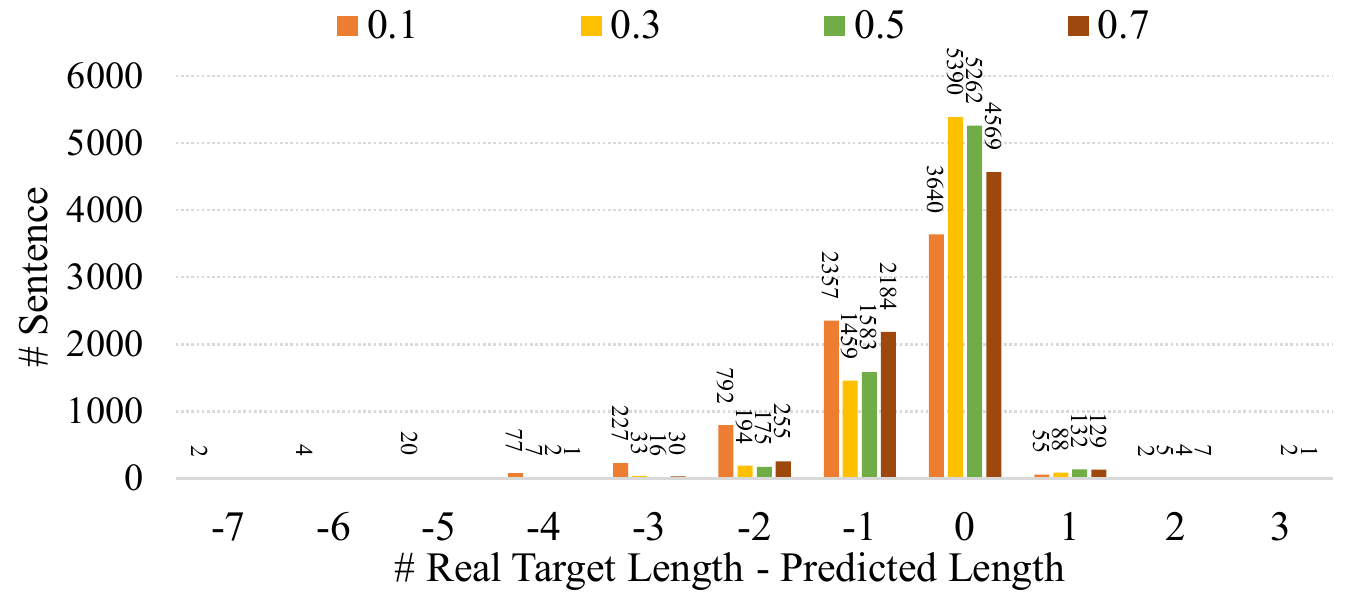}
	\caption{The analysis of the predicted length. The histogram shows the difference between the target length and the predicted length.}
\end{figure}
\vspace{-10pt}

\begin{figure}[t]
	\centering
	\subfigure[The realtionship bettween trigger and word boundaries]{
	\label{fig:ctc}
	\includegraphics[width=\linewidth]{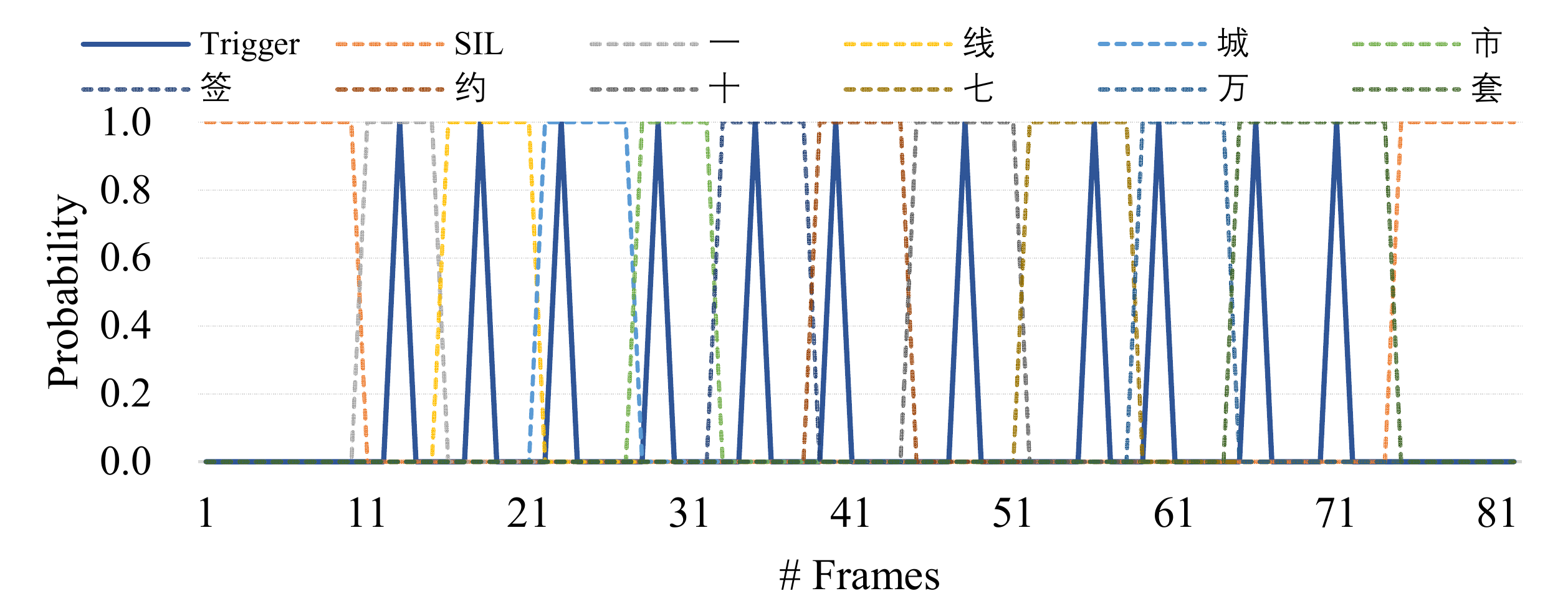}
}

	\subfigure[Attention mechanism visualization]{
	\label{fig:attn}
	\includegraphics[width=\linewidth]{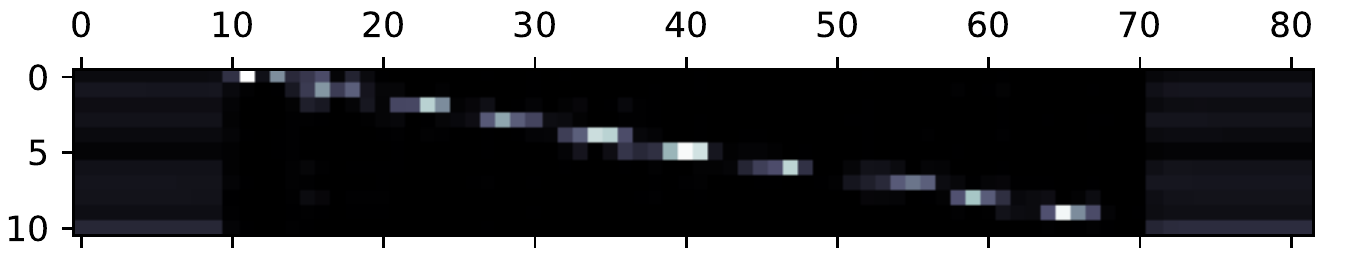}
}
	\caption{We visually analyzed the test set sentences 'BAC009S0764W0149'. (a)The line chart shows the relationship between trigger position and character pronunciation boundaries. The dotted lines indicate the pronunciation boundary information and the spikes present the spike-like posterior probability of the CTC module. (b) is from the 4th source attention mechanism of the decoder.}
	\vspace{-10pt}
\end{figure}


\begin{table}[t]
	\caption{Compare with other models in performance and real-time factor.}
	\label{tab:models}
	\centering
	\begin{threeparttable}
	\begin{tabular}{c|ccc}
		\toprule
		\textbf{Model} & \textbf{DEV} & \textbf{TEST} & \textbf{RTF} \\
		\hline
		TDNN-Chain (Kaldi) \cite{povey2016purely} & - & 7.45 & -  \\
		LAS\cite{8682490} & - & 10.56 & - \\
		Speech-Transformer * & 6.57 & 7.37 & 0.0504   \\
		SA-Transducer $\dagger$ \cite{Tian2019} & 8.30 & 9.30 & 0.1536  \\
		SAN-CTC * \cite{salazar2019self} & 7.83 & 8.74 & 0.0168  \\
		Sync-Transformer $\dagger$ \cite{tian2019synchronous} & 7.91 & 8.91 & 0.1183  \\
		NAT-MASKED * \cite{chen2019non} & 7.16 & 8.03 & 0.0058  \\
		ST-NAT(ours) & 6.88 & 7.67 & \textbf{0.0056} \\
		ST-NAT+LM(ours) & \textbf{6.39} & \textbf{7.02} & 0.0292  \\
		\bottomrule
	\end{tabular}
   	\begin{tablenotes}
		\item[*] These models are re-implemented by ourselves according to the papers.
		\item[$\dagger$] We supplement the RTF of our previous two models.
	\end{tablenotes}
	\end{threeparttable}
\vspace{-5pt}
\end{table}

\subsubsection{Compare with other models.}

We also compare our proposed ST-NAT model with various main-stream models, e.g. traditional model, CTC-based model, transducer model, and attention-based sequence-to-sequence model. Under the same training condition and the same model parameters, we train a Speech-Transformer\cite{dong2018speech}, NAT-MASKED \cite{chen2019non}, and our proposed ST-NAT model, where the speech transformer applies a beam search with beam width 5 to decoding utterances. 

From Table.\ref{tab:models}, we can find the ST-NAT models can achieve comparable performance with the advanced speech-transformer model \cite{dong2018speech} and TDNN-Chain model \cite{povey2016purely}, which is better than LAS. From another perspective, the ST-NAT has the fastest inference speed among them, which is only about 1/10 of speech-transformer. The ST-NAT with a transformer language model can achieve the best CER of 7.02\% on the test set and an RTF of 0.0292. 

Compared with the streaming end-to-end model, e.g. SAN-CTC \cite{salazar2019self}, Sync-Transformer \cite{tian2019synchronous}, and SA-Transducer \cite{Tian2019}, the ST-NAT can not only achieve the best performance, but also the fastest inference speed. We suppose that the ST-NAT can decode an utterance with all context and without temporal dependencies.  

By contrast, we also re-implement a NAT-MASKED model in a BERT-like way \cite{chen2019non}, which adopts a fixed-length (set as 60) mask sequence as the input. The NAT-MASKED has the same parameters as our ST-NAT except for the CTC module. We find the ST-NAT can achieve better performance. We guess that it is difficult for the model to learn to predict the target words(or characters) and the target length jointly. Both of them have a very close inference speed.

\section{Conclusions and Future Works}

To estimate the length of the target sequence accurately and accelerate the inference speech, we proposed a spike-triggered non-autoregressive transformer (ST-NAT) for end-to-end speech recognition, which introduce a CTC module to predict the target length and accelerate the convergence. The ST-NAT adopts the encode states corresponding to the positions of spikes as the input of the decoder. In the inference process, ST-NAT can count the number of spikes to avoid redundant calculations. We conduct all experiments on a public Chinese mandarin dataset AISEHLL-1. The results show that the CTC module can accurately predict the length of the target sequence. The ST-NAT model has achieved achieve comparable performance with the advanced speech transformer model. However, the ST-NAT has a real-time factor of 0.0056, which exceeds all mainstream models.  What's more, the ST-NAT with a language model can still have a very high inference speed. In the future, we will try to utilize the CTC module for joint decoding to improve the performance of the model during inference.


\bibliographystyle{IEEEtran}

\bibliography{mybib}

\begin{thebibliography}{10}
\providecommand{\url}[1]{#1}
\csname url@samestyle\endcsname
\providecommand{\newblock}{\relax}
\providecommand{\bibinfo}[2]{#2}
\providecommand{\BIBentrySTDinterwordspacing}{\spaceskip=0pt\relax}
\providecommand{\BIBentryALTinterwordstretchfactor}{4}
\providecommand{\BIBentryALTinterwordspacing}{\spaceskip=\fontdimen2\font plus
\BIBentryALTinterwordstretchfactor\fontdimen3\font minus
  \fontdimen4\font\relax}
\providecommand{\BIBforeignlanguage}[2]{{%
\expandafter\ifx\csname l@#1\endcsname\relax
\typeout{** WARNING: IEEEtran.bst: No hyphenation pattern has been}%
\typeout{** loaded for the language `#1'. Using the pattern for}%
\typeout{** the default language instead.}%
\else
\language=\csname l@#1\endcsname
\fi
#2}}
\providecommand{\BIBdecl}{\relax}
\BIBdecl

\bibitem{bahdanau2014neural}
D.~Bahdanau, K.~Cho, and Y.~Bengio, ``Neural machine translation by jointly
  learning to align and translate,'' \emph{arXiv preprint arXiv:1409.0473},
  2014.

\bibitem{chorowski2015attention}
C.~J. K, B.~Dzmitry, S.~Dmitriy, C.~Kyunghyun, and B.~Yoshua, ``Attention-based
  models for speech recognition,'' in \emph{Advances in neural information
  processing systems}, 2015, pp. 577--585.

\bibitem{chan2016listen}
C.~William, J.~Navdeep, L.~Quoc, and V.~Oriol, ``Listen, attend and spell: A
  neural network for large vocabulary conversational speech recognition,'' in
  \emph{2016 IEEE International Conference on Acoustics, Speech and Signal
  Processing (ICASSP)}.\hskip 1em plus 0.5em minus 0.4em\relax IEEE, 2016, pp.
  4960--4964.

\bibitem{vaswani2017attention}
A.~Vaswani, N.~Shazeer, N.~Parmar, J.~Uszkoreit, L.~Jones, A.~N. Gomez,
  {\L}.~Kaiser, and I.~Polosukhin, ``Attention is all you need,'' in
  \emph{Advances in neural information processing systems}, 2017, pp.
  5998--6008.

\bibitem{kim2017joint}
K.~Suyoun, H.~Takaaki, and W.~Shinji, ``Joint ctc-attention based end-to-end
  speech recognition using multi-task learning,'' in \emph{2017 IEEE
  international conference on acoustics, speech and signal processing
  (ICASSP)}.\hskip 1em plus 0.5em minus 0.4em\relax IEEE, 2017, pp. 4835--4839.

\bibitem{dong2018speech}
D.~Linhao, X.~Shuang, and X.~Bo, ``Speech-transformer: a no-recurrence
  sequence-to-sequence model for speech recognition,'' in \emph{2018 IEEE
  International Conference on Acoustics, Speech and Signal Processing
  (ICASSP)}.\hskip 1em plus 0.5em minus 0.4em\relax IEEE, 2018, pp. 5884--5888.

\bibitem{lee2018deterministic}
J.~Lee, E.~Mansimov, and K.~Cho, ``Deterministic non-autoregressive neural
  sequence modeling by iterative refinement,'' \emph{arXiv preprint
  arXiv:1802.06901}, 2018.

\bibitem{gu2017non}
J.~Gu, J.~Bradbury, C.~Xiong, V.~O. Li, and R.~Socher, ``Non-autoregressive
  neural machine translation,'' \emph{arXiv preprint arXiv:1711.02281}, 2017.

\bibitem{ma2019flowseq}
X.~Ma, C.~Zhou, X.~Li, G.~Neubig, and E.~Hovy, ``Flowseq: Non-autoregressive
  conditional sequence generation with generative flow,'' \emph{arXiv preprint
  arXiv:1909.02480}, 2019.

\bibitem{wang2019non}
Y.~Wang, F.~Tian, D.~He, T.~Qin, C.~Zhai, and T.-Y. Liu, ``Non-autoregressive
  machine translation with auxiliary regularization,'' in \emph{Proceedings of
  the AAAI Conference on Artificial Intelligence}, vol.~33, 2019, pp.
  5377--5384.

\bibitem{chen2019non}
N.~Chen, S.~Watanabe, J.~Villalba, and N.~Dehak, ``Non-autoregressive
  transformer automatic speech recognition,'' \emph{arXiv preprint
  arXiv:1911.04908}, 2019.

\bibitem{libovicky2018end}
J.~Libovick{\`y} and J.~Helcl, ``End-to-end non-autoregressive neural machine
  translation with connectionist temporal classification,'' \emph{arXiv
  preprint arXiv:1811.04719}, 2018.

\bibitem{moritz2019triggered}
N.~Moritz, T.~Hori, and J.~Le~Roux, ``Triggered attention for end-to-end speech
  recognition,'' in \emph{ICASSP 2019 IEEE International Conference on
  Acoustics, Speech and Signal Processing (ICASSP)}.\hskip 1em plus 0.5em minus
  0.4em\relax IEEE, 2019, pp. 5666--5670.

\bibitem{moritz2019streaming}
------, ``Streaming end-to-end speech recognition with joint ctc-attention
  based models,'' in \emph{Proc. IEEE Workshop on Automatic Speech Recognition
  and Understanding (ASRU)}, 2019.

\bibitem{dauphin2017language}
Y.~N. Dauphin, A.~Fan, M.~Auli, and D.~Grangier, ``Language modeling with gated
  convolutional networks,'' in \emph{Proceedings of the 34th International
  Conference on Machine Learning-Volume 70}.\hskip 1em plus 0.5em minus
  0.4em\relax JMLR. org, 2017, pp. 933--941.

\bibitem{Tian2019}
Z.~Tian, J.~Yi, J.~Tao, Y.~Bai, and Z.~Wen, ``{Self-Attention Transducers for
  End-to-End Speech Recognition},'' in \emph{Proc. Interspeech 2019}, 2019, pp.
  4395--4399.

\bibitem{fan2019speaker}
Z.~Fan, J.~Li, S.~Zhou, and B.~Xu, ``Speaker-aware speech-transformer,'' in
  \emph{2019 IEEE Automatic Speech Recognition and Understanding Workshop
  (ASRU)}.\hskip 1em plus 0.5em minus 0.4em\relax IEEE, 2019, pp. 222--229.

\bibitem{graves2006connectionist}
A.~Graves, S.~Fern{\'a}ndez, F.~Gomez, and J.~Schmidhuber, ``Connectionist
  temporal classification: labelling unsegmented sequence data with recurrent
  neural networks,'' in \emph{Proceedings of the 23rd international conference
  on Machine learning}, 2006, pp. 369--376.

\bibitem{de2005tutorial}
P.-T. De~Boer, D.~P. Kroese, S.~Mannor, and R.~Y. Rubinstein, ``A tutorial on
  the cross-entropy method,'' \emph{Annals of operations research}, vol. 134,
  no.~1, pp. 19--67, 2005.

\bibitem{park2019specaugment}
D.~S. Park, W.~Chan, Y.~Zhang, C.-C. Chiu, B.~Zoph, E.~D. Cubuk, and Q.~V. Le,
  ``Specaugment: A simple data augmentation method for automatic speech
  recognition,'' \emph{arXiv preprint arXiv:1904.08779}, 2019.

\bibitem{povey2016purely}
D.~Povey, V.~Peddinti, D.~Galvez, P.~Ghahremani, V.~Manohar, X.~Na, Y.~Wang,
  and S.~Khudanpur, ``Purely sequence-trained neural networks for asr based on
  lattice-free mmi.'' in \emph{Interspeech}, 2016, pp. 2751--2755.

\bibitem{8682490}
C.~{Shan}, C.~{Weng}, G.~{Wang}, D.~{Su}, M.~{Luo}, D.~{Yu}, and L.~{Xie},
  ``Component fusion: Learning replaceable language model component for
  end-to-end speech recognition system,'' in \emph{ICASSP 2019 - 2019 IEEE
  International Conference on Acoustics, Speech and Signal Processing
  (ICASSP)}, 2019, pp. 5361--5635.

\bibitem{salazar2019self}
J.~Salazar, K.~Kirchhoff, and Z.~Huang, ``Self-attention networks for
  connectionist temporal classification in speech recognition,'' in
  \emph{ICASSP 2019-2019 IEEE International Conference on Acoustics, Speech and
  Signal Processing (ICASSP)}.\hskip 1em plus 0.5em minus 0.4em\relax IEEE,
  2019, pp. 7115--7119.

\bibitem{tian2019synchronous}
Z.~Tian, J.~Yi, Y.~Bai, J.~Tao, S.~Zhang, and Z.~Wen, ``Synchronous
  transformers for end-to-end speech recognition,'' in \emph{ICASSP 2020 IEEE
  International Conference on Acoustics, Speech and Signal Processing
  (ICASSP)}.\hskip 1em plus 0.5em minus 0.4em\relax IEEE, 2020, pp. 5666--5670.

\end{thebibliography}

\end{document}